\documentclass[11pt,twoside]{article}

\usepackage{asp2006}
\usepackage{epsf}
\usepackage{psfig}
\usepackage{lscape}

\begin{document}

\title{The {\it Spitzer} 24\micron\ Photometric Light Curve of
the Eclipsing M-dwarf Binary GU Bo\"otis} 


\author{Kaspar von Braun, Gerard T. van Belle, David Ciardi}  
\affil{Michelson Science Center, California Institute of
Technology, MC 100-22, Pasadena, CA 91125; kaspar, gerard,
ciardi@ipac.caltech.edu} 

\author{Mercedes L\'{o}pez-Morales}
\affil{Carnegie Fellow, Department of Terrestrial Magnetism, 
Carnegie Institution of Washington, 5241 Broad Branch Rd. NW,
Washington, DC 20015; mercedes@dtm.ciw.edu}

\author{D. W. Hoard, and Stefanie Wachter}
\affil{Spitzer Science Center, California Institute of
Technology, MC 220-6, Pasadena, CA 91125;
hoard, wachter@ipac.caltech.edu}


\begin{abstract}
We present a carefully controlled set of Spitzer 24 \micron\ MIPS time series
observations of the low mass eclipsing binary star GU Bo\"otis (GU Boo).  Our
data cover three secondary eclipses of the system: two consecutive events and
an additional eclipse six weeks later. The study's main purpose is the long
wavelength characterization of GU Boo's light curve, independent of limb
darkening and less sensitive to surface features such as spots.  Its analysis
allows for independent verification of the results of optical studies of GU
Boo. Our mid-infrared results show good agreement with previously obtained
system parameters. In addition, the analysis of light curves of other objects
in the field of view serves to characterize the photometric stability and
repeatability of {\it Spitzer's} MIPS-24 at flux densities between
approximately 300--2,000$\mu$Jy. We find that the light curve root mean square
about the median level falls into the 1--4\% range for flux densities higher
than 1 mJy.

\end{abstract}


\section{Why is GU Bo\"otis Important?}\label{introduction}

GU Bo\"otis is a nearby, low-mass detached eclipsing binary system, consisting
of two nearly equal mass M-dwarfs \citep{lmr05}. It is one of currently very
few known nearby ($<$ 200 pc) double-lined, detached eclipsing binary (DEB)
systems composed of two low-mass companions \citep{lm07}.  Eclipsing binaries
can be used as tools to constrain fundamental stellar properties such as mass,
linear radius, and effective temperature. Given the fact that over 70\% of the
stars in the Milky Way are low-mass objects with $M< 1 M_\odot$ \citep{hik97},
coupled with the considerable uncertainty over the mass-radius relation for
low-mass stars, objects such a GU Bo\"otis are of particular interest in
exploring the low-mass end of the Hertzsprung-Russell Diagram.

The characterization of the effects of limb darkening and star spots
introduces free parameters and thus statistical uncertainty in the
calculation of the stellar radii and masses.  Using the {\it Spitzer Space
Telescope}, we obtained 24 $\mu$m time series observations of three separate
instances of GU Boo's secondary eclipse (see Table 1) to create
a light curve far enough in the infrared to not be contaminated by the effects
of limb darkening and star spots. 

A further goal of our study is to characterize the photometric stability of
the Multiband Imaging Photometer (MIPS) on {\it Spitzer} at 24 $\mu$m over
short and long time scales. Time-series observing is atypical (albeit
increasingly common) for {\it Spitzer} which is the reason why there are very
few published photometric light curves based on {\it Spitzer}
observations. The recent spectacular observations of primary and secondary
eclipses of transiting planets are notable exception \citep[see for
instance][]{cam05,dsr05}. Of these, the \citet{dsr05} study was performed at
24 $\mu$m. We therefore observed two consecutive secondary eclipses of GU Boo
about 12 hours apart (observing sets 1 and 2), and then a third event about
six weeks later (observing set 3). Table 1 gives an overview. 

We describe our MIPS-24 observations and data reduction procedure in \S2,
present our results concerning GU Boo's light curve and the photometric
stability of {\it Spitzer} and MIPS-24 in \S3, and summarize in \S4.


\section{Observations and Data Reduction}\label{reductions}

We used MIPS-24 aboard the {\it Spitzer Space Telescope} \citep{wrl04} to
observe GU Bo\"otis in February and April of 2006, as outlined in Table 1.
MIPS-24, the 24 $\mu$m array, is a Si:As detector with 128 $\times$ 128
pixels, an image scale of 2.55" pixel$^{-1}$, and a field of view of 5.4'
$\times$ 5.4' \citep{rye04}.  Our exposures were obtained using the standard
MIPS 24 $\mu$m small field photometry pattern \citep[for details, see, for
instance,][]{rhs06}.

Our goal was to observe three independent secondary eclipses of GU Boo: two
consecutive ones and another one several weeks after the first two.  Of our
total of nine of {\it Spitzer's} Astronomical Observation Requests (AORs),
three were used for each secondary eclipse event (see Table 1).  Each AOR
contained eight observing cycles with 36 individual exposures each. The first
exposure in each cycle is 9s long, the subsequent 35 are 10s long. The first
two exposures of every cycle were discarded due to a ``first frames
effect''. This procedure left 34 frames per cycle, 272 frames per AOR, 816
frames per secondary eclipse event, and 2448 frames for the entire project
(all 10s exposure time)\footnote{For background information on the {\it
Spitzer} and MIPS operations, we refer the reader to the Spitzer Observer's
Manual (SOM -- http://ssc.spitzer.caltech.edu/documents/som/). For information
specifically related to MIPS data reduction, please consult the MIPS Data
Handbook (MDH -- http://ssc.spitzer.caltech.edu/mips/dh/) and \citet{gre05}.}.


\begin{table}[!ht]
\caption{Spitzer MIPS-24 observations of GU Bo\"otis}
\smallskip
\begin{center}
{\small
\begin{tabular}{ccccc}
\tableline
\noalign{\smallskip}
Date (2006) & MIPS Campaign & Obs. Set & AORs & Exposures\tablenotemark{a} \\
\noalign{\smallskip}
\tableline
\noalign{\smallskip}
Feb 20 &   {\it MIPS006500} & 1 & {\it 16105472} &        860 \\
           &     {\it } & & {\it 16105216} &            \\
           &     {\it } & & {\it 16104960} &            \\
Feb 21 &   {\it MIPS006500} & 2 & {\it 16104704} &        860 \\
           &     {\it } & & {\it 16104448} &            \\
           &     {\it } & & {\it 16104192} &            \\
Apr 01 &   {\it MIPS006700} & 3 & {\it 16103936} &        860 \\
           &     {\it } & & {\it 16103680} &            \\
           &     {\it } & & {\it 16103424} &            \\
\noalign{\smallskip}
\tableline
\end{tabular}
\tablenotetext{a}{10 seconds per exposure.}
}
\end{center}
\end{table}


The MIPS-24 data are provided by the {\it Spitzer Archive} in the
(flatfielded) Basic Calibrated Data (BCD) format. We applied further
post-processing to these data in order to correct for small scale artifacts,
in particular using IRAF's\footnote{IRAF is distributed by the National
Optical Astronomy Observatory, which is operated by the Association of
Universities for Research in Astronomy, Inc, under cooperative agreement with
the National Science Foundation.} CCDRED package to remove the weak
``jailbar'' features in the images (as described in the MDH).

\begin{figure}
\plotone{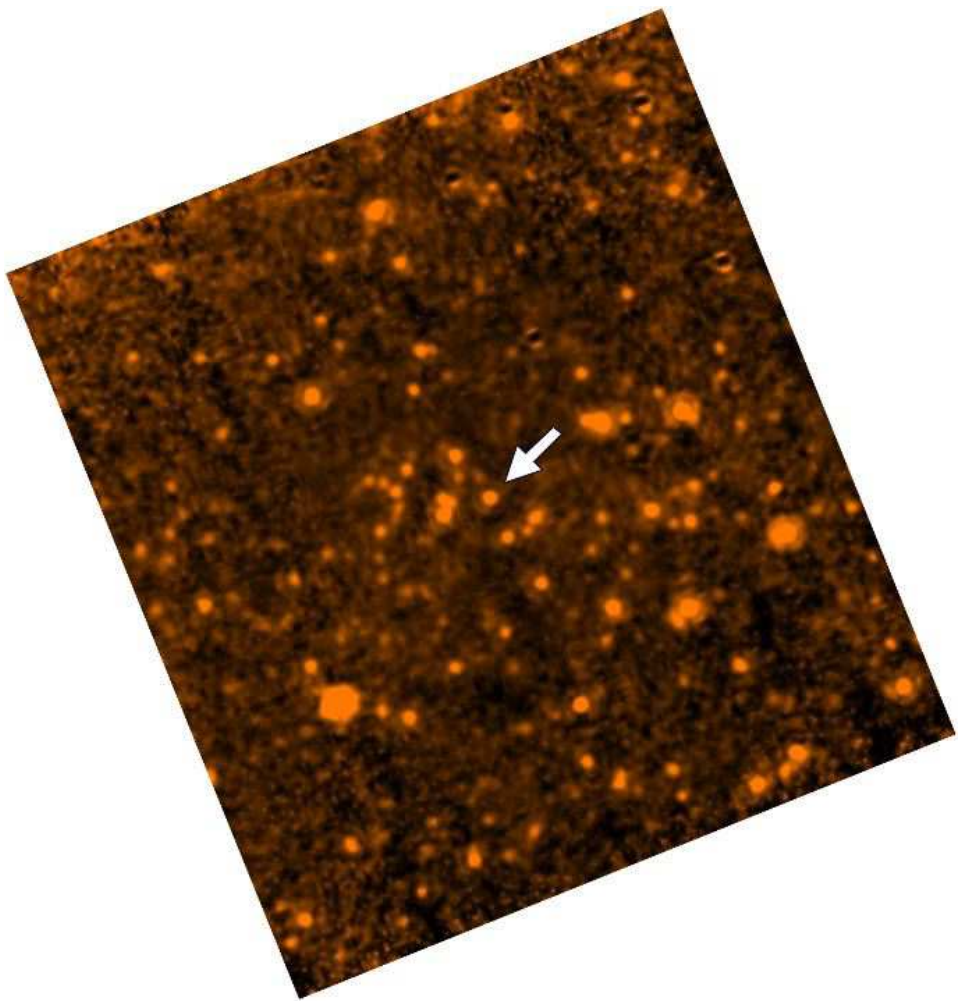} 
\caption{\label{mosaic_AOR} A {\it Spitzer} MIPS 24 $\mu$m mosaic of 
GU Bo\"otis (marked with arrow at the center of the image). This mosaic was
created using all 272 frames in one AOR and is about 5 arcmin on the
side. North is up, east is to the left.}
\end{figure}

The {\it Spitzer} software package \textsf{mopex} \citep{mk05,mm05} was used
for co-adding the individual MIPS frames into mosaics of 17 frames, using
overlap correction and outlier rejection in the process. The choice of 17
frames was made (1) to obtain a high signal-to-noise ratio (SNR) for a
measured stellar flux density in a resulting combined image and subsequent
data point in the respective star's light curves, (2) to maintain a
sufficiently high effective observing cadence to temporally resolve elements
of GU Boo's light curve, and (3) not to be forced to combine frames from
different cycles into a single light curve data point (see Table 1). The
interpolated, remapped mosaics have a pixel scale of 2.45'' pixel$^{-1}$. We
show in Figure \ref{mosaic_AOR} the MIPS-24 field of view of GU Bo\"otis.


For photometric reductions of the mosaiced images, we utilized the
\textsf{apex} component of \textsf{mopex} to perform point-source extraction
as described in \citet{mm05}\footnote{Also see information on \textsf{apex} at
http://ssc.spitzer.caltech.edu/postbcd/apex.html and the User's Guide at
http://ssc.spitzer.caltech.edu/postbcd/doc/apex.pdf}. This step included
background subtraction of the images, and the fitting of a resampled point
response function (PRF), derived from our own data. In order to match the PRF
centroid as closely as possible to the centroid of the stellar profile, the
first Airy ring was initially subtracted from the stellar profile, and
detection was then performed on the resulting image.


\section{Results}\label{Results}


\subsection{Analysis of GU Boo's Light Curve}

\begin{figure}
\plotone{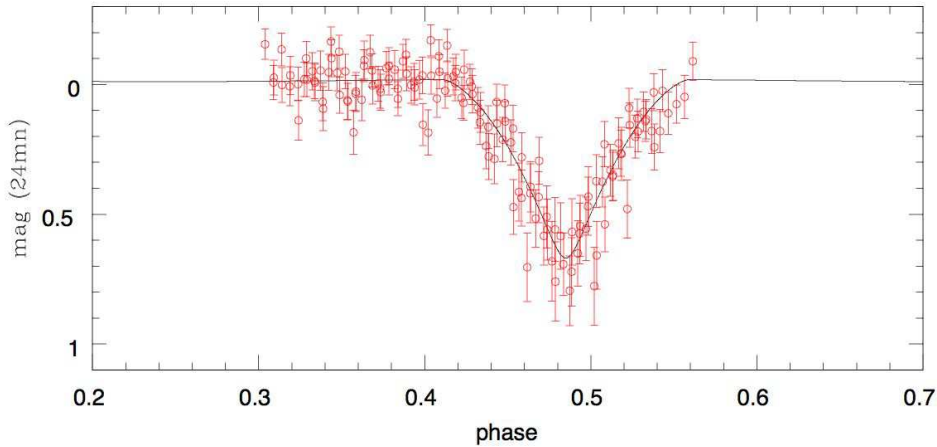} 
\caption{\label{lightcurve} Folded 24 \micron\ light curve for all 3 observed
secondary eclipses of GU Boo. The preliminary fit is overlaid. The ordinate is
scaled to 24 \micron\ magnitudes with the zero point corresponding to the
out-of-eclipse flux density level.}
\end{figure}

We show in Fig. \ref{lightcurve} the phased light curve of GU Boo. 
Based on preliminary light curve fitting to the relative flux density levels
(scaled to magnitudes) shown in Figure \ref{lightcurve}, we find that our
results are consistent with the system parameters derived from the optical
study of the system in \citet{lmr05}.  The orbital period and initial epoch of
the primary eclipse were set to the values given in the ephemerides equation
derived by \citet{lmr05}. We further fixed the mass ratio and the radius ratio
of the stars, as well as the eccentricity of the system ($e$=0) to the values
obtained in that work. We assumed no limb darkening effects in the light
curves, as expected for observations that far into the infrared \citep[][and
references therein]{cdg95,rhs06,cbb07}, and no significant gravitational
darkening or reflection effects, based on the spherical shape of the stars and
the similarity in effective temperatures. All these are reasonable
assumptions, based on the results of the study of GU Boo at visible
wavelengths, and they are in fact hard to test in detail, given the
photometric precision of the {\it Spitzer} light curve at this flux density
level. Table 2 gives our estimates of GU Boo's system parameters.


\begin{table}[!ht]
\caption{GU Boo System Parameters}
\smallskip
\begin{center}
{\small
\begin{tabular}{lr}
\tableline
\noalign{\smallskip}
Parameter & Value \\
\noalign{\smallskip}
\tableline
\noalign{\smallskip}

Orbital Period (days)\tablenotemark{a} & $0.488728  \pm  0.000002 $ \\
Orbital Eccentricity\tablenotemark{a} &          0 (fixed)\\
Mass Ratio (M2/M1) \tablenotemark{a} & $0.9832  \pm  0.0069 $ \\
Radius of Secondary Component ($R_{\sun}$) & 0.66 $\pm$ 0.02 (0.62\tablenotemark{a})\\
Orbital Inclination $i$ (degrees) & 89.3 $\pm$ 0.8 (87.6\tablenotemark{a})\\
\noalign{\smallskip}
\tableline
\end{tabular}
\tablenotetext{a}{From \citet{lmr05}.}
}
\end{center}
\end{table}


\subsection{MIPS-24 Photometry Stability}

\begin{figure}
\plotone{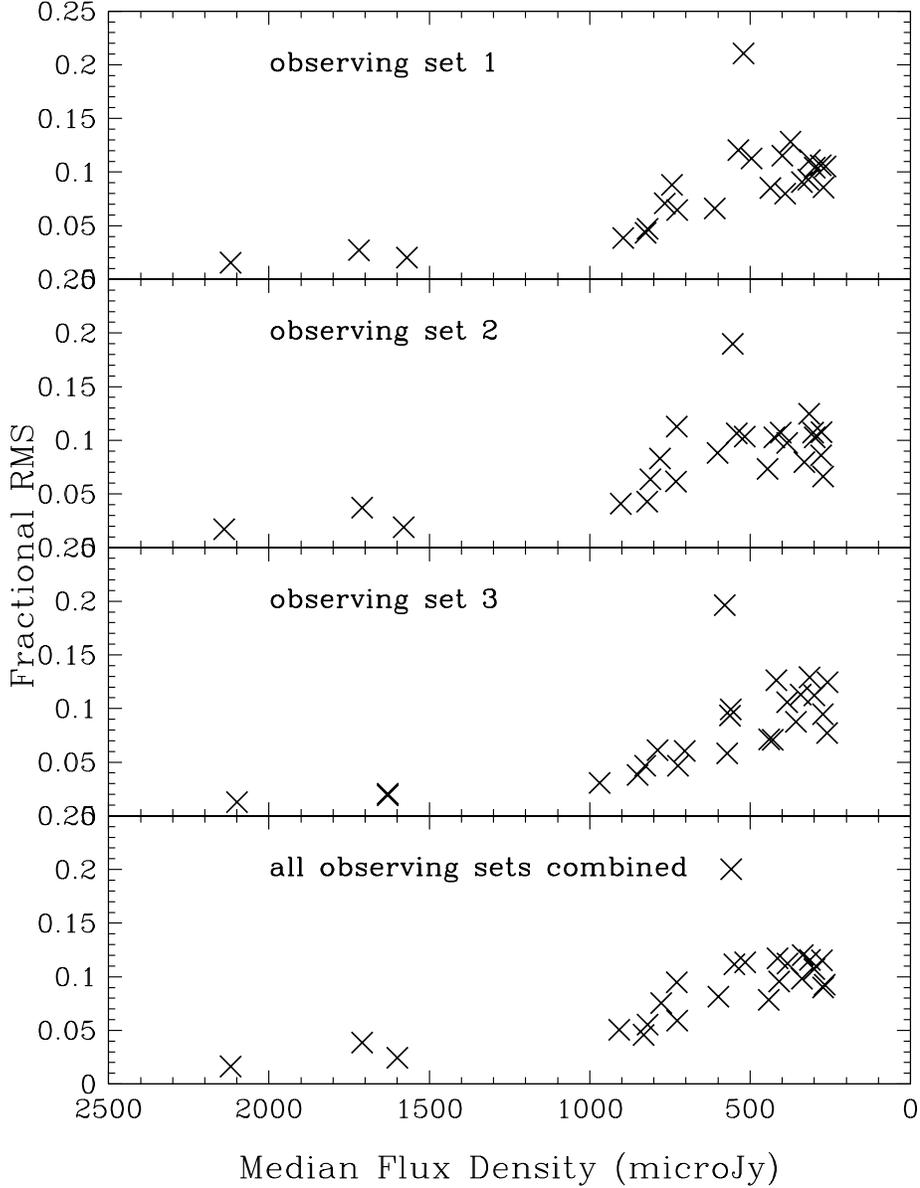} 
\caption{\label{global_rms} 
A plot of median flux density versus fractional rms for the 24 stars that have
photometry for more than 72 out of 144 observational epochs. Shown are the
individual 3 observing sets (see Table 1) to illustrate the repeatability of
{\it Spitzer}/MIPS-24 within individual observing sets, as well as a plot of
all 3 sets combined (to show the inter-set stability). The data point with the
highest fractional rms is GU Boo. }
\end{figure}

Figure \ref{global_rms} shows the fractional rms around median values for all
objects with more than 72 out of a total of 144 observational epochs for each
individual observing set as well as for the three sets combined. Observing
sets 1 and 2 were obtained during the MIPS006500 campaign, observing set 3
during MIPS006700 (Table 1). We find that inter-set repeatability of {\it
Spitzer's} MIPS-24 is comparable to its repeatability within a set, both in
terms of median flux density level as well as rms values. For the objects with
a flux density in excess of 1 mJy, the rms values approach the 1--4 \%
level. The light curves of all objects in the field (other than GU Boo itself)
are flat with different amounts of random scatter around the median flux
density level.


\section{Summary}\label{summary}

We used MIPS-24 onboard the {\it Spitzer Space Telescope} to obtain
time-series photometry of the M-dwarf DEB GU Boo. Our observations cover three
secondary eclipse events, two consecutive ones and a further event six weeks
later.  Our mid-IR analysis of GU Boo's light curve is less affected by
stellar surface features than its optical counterpart. The results
show good agreement with the previously obtained system parameters based on
optical and near-IR work.  Finally, we find that the repeatability of MIPS-24
photometry is consistent over all temporal scales we sampled: within an
observing set and on time scales of 24 hours and six weeks.


\acknowledgements 

We gratefully acknowledge the availability of {\it Spitzer} Director's
Discretionary Time (DDT) for this project. We furthermore thank D. Frayer,
S. Carey, J. Colbert, and P. Lowrance for their help with the MIPS data
reduction pipeline.




\begin{thebibliography}{}
\bibitem[Charbonneau et al.(2005)]{cam05} Charbonneau, D., et al.\ 2005, \apj,
626, 523
\bibitem[Ciardi et al.(2007)]{cbb07} Ciardi, D.~R., et al.\ 
2007, \apj, 659, 1623 
\bibitem[Claret et al.(1995)]{cdg95} Claret, A., 
Diaz-Cordoves, J., \& Gimenez, A.\ 1995, \aaps, 114, 247 
\bibitem[Deming et al.(2005)]{dsr05} Deming, D., Seager, S., Richardson,
L.~J., \& Harrington, J.\ 2005, \nat, 434, 740
\bibitem[Gordon et al.(2005)]{gre05} Gordon, K.~D., et al.\ 
2005, \pasp, 117, 503 
\bibitem[Henry et al.(1997)]{hik97} Henry, T.~J., Ianna, P.~A., Kirkpatrick,
J.~D., \& Jahreiss, H.\ 1997, \aj, 114, 388 
\bibitem[Makovoz \& Khan(2005)]{mk05} Makovoz, D., \& Khan, 
I.\ 2005, Astronomical Data Analysis Software and Systems XIV, 347, 81 
\bibitem[Makovoz \& Marleau(2005)]{mm05} Makovoz, D., \& 
Marleau, F.~R.\ 2005, \pasp, 117, 1113 
\bibitem[L{\'o}pez-Morales \& Ribas(2005)]{lmr05} L{\'o}pez-Morales, M., \& Ribas, I.\ 2005, \apj, 631, 1120 
\bibitem[L{\'o}pez-Morales(2007)]{lm07} L{\'o}pez-Morales, M.\ 2007, \apj, 660, 732
\bibitem[Richardson et al.(2006)]{rhs06} Richardson, L.~J., Harrington, J.,
Seager, S., \& Deming, D.\ 2006, \apj, 649, 1043
\bibitem[Rieke et al.(2004)]{rye04} Rieke, G.~H., et al.\ 
2004, \apjs, 154, 25  
\bibitem[Werner et al.(2004)]{wrl04} Werner, M.~W., et al.\ 
2004, \apjs, 154, 1 
\end{thebibliography}
\end{document}